\documentclass[sn-mathphys]{sn-jnl}

\usepackage{bm}

\jyear{2022}%
\begin{document}

\title[Generation of Turbulent States]{Generation of Turbulent States
using Physics-Informed Neural Networks}

\author[1,2]{\fnm{Sofía} \sur{Angriman}}\email{sangriman@df.uba.ar}
\author[1,2]{\fnm{Pablo} \sur{Cobelli}}\email{cobelli@df.uba.ar}
\author[1,2]{\fnm{Pablo D.} \sur{Mininni}}\email{mininni@df.uba.ar}
\author[3]{\fnm{Martín} \sur{Obligado}}\email{Martin.Obligado@univ-grenoble-alpes.fr}
\author*[4]{\fnm{Patricio} \sur{Clark Di Leoni}}\email{pclarkdileoni@udesa.edu.ar}

\affil[1]{\orgname{Universidad de Buenos Aires}, \orgdiv{Facultad de Ciencias Exactas y Naturales, Departamento de Física}, \orgaddress{\street{Ciudad Universitaria}, \city{Buenos Aires}, \postcode{1428}, \country{Argentina}}}

\affil[2]{\orgname{CONICET - Universidad de Buenos Aires}, \orgdiv{Instituto de F\'{\i}sica del Plasma (INFIP)}, \orgaddress{\street{Ciudad Universitaria}, \city{Buenos Aires}, \postcode{1428}, \country{Argentina}}}

\affil[3]{\orgname{Universit\'e Grenoble Alpes, CNRS}, \orgdiv{Grenoble-INP, LEGI}, 
\orgaddress{\postcode{F-38000}, \city{Grenoble}, \country{France}}}
\affil[4]{\orgdiv{Departmento de Ingeniería}, \orgname{Universidad de San Andrés},
\orgaddress{\city{Buenos Aires}, 
\country{Argentina}}}

\abstract{When modelling turbulent flows, it is often the
case that information on the forcing terms or the boundary
conditions is either not available or overly complicated and 
expensive to implement. Instead, some flow features, such as 
the mean velocity profile or its statistical moments, may be
accessible through experiments or observations. We present a 
method based on physics-informed neural networks to generate 
turbulent states subject to a set of given conditions. The 
physics-informed method ensures the final state approximates a
valid flow. We show examples of different statistical conditions
that can be used to prepare states, motivated by experimental 
and atmospheric problems. Lastly, we show two ways of scaling 
the resolution of the prepared states. One is through the use of 
multiple and parallel neural networks. The other uses nudging, a
synchronization-based data assimilation technique that leverages
the power of specialized numerical solvers.  }


\maketitle

\section{Introduction}
\label{introduction}

Generating turbulent states can be a clear but highly computationally intensive task, in the best of cases, or a dauntingly complex task, in the worst ones.
Modern computational capabilities, coupled with the advent of sophisticated parallel methods \cite{mininni_hybrid_2011}, have made Direct Numerical Simulations (DNSs) broadly available at a wide range of resolutions, but they still remain prohibitively expensive for many applications. The problem becomes even more complex when the exact form of the forcing terms or boundary conditions in the system of interest is not known.
Techniques for generating synthetic homogeneous and isotropic turbulent states based on Fourier decompositions go back as far as Kraichnan \cite{kraichnan_diffusion_1970} and have been extended to wall-bounded flows, where the goal is usually to generate inflow conditions for boundary layer or channel flow simulations \cite{dhamankar_overview_2015, wu_inflow_2017}. The idea behind these studies is to generate turbulent states by superposing Fourier modes with random phases, but setting their amplitudes and temporal evolution so that they satisfy Kolmogorov's spectra. Another condition that is often imposed is that the resulting fields should be divergence-free. 
In atmospheric and oceanic sciences, the problem of generating turbulent states also appears when trying to prepare initial conditions for a simulation. The normal course of action in those cases is to merge the result of a previous simulation with current observations, in a process known as Data Assimilation \cite{kalnay_atmospheric_2003}. Data Assimilation (DA) can help in correcting trajectories in phase space resulting from uncertainties in the initial conditions, as well as in incorporating effects coming from missing or incorrectly modeled physics. While variational and ensemble-based methods are the two main families of DA algorithms, stochastic methods \cite{Cotter2020} and machine and deep learning techniques \cite{chantry_opportunities_2021} have been making great strides in this area.

The DA approaches mentioned above work, for the most part, with direct data, as they can incorporate, for example, pressure measurements at a particular time and place.
Sometimes though, one only wants to assimilate statistical properties of a flow. This is the case, e.g., when correcting predictions stemming from a turbulence model \cite{foures_data-assimilation_2014, mons_ensemble-variational_2021}, in which case one seeks to assimilate the correct mean velocity or another statistical moment. Another prime example, and the one which acts as motivation for this work, is when performing asynchronous point-wise measurements at different locations of a flow. While each measurement might be time resolved and may correspond to a given realization of the flow, the different measurements cannot be put together to reconstruct the whole instantaneous flow, and thus only statistical information is available.

The purpose of the present work is to address such problems. We present a method to prepare turbulent states subject to a given set of conditions. These conditions can be, for example, the shape of the mean velocity profile, or the value of high order statistical moments. The method takes a seed field, e.g., coming from a homogeneous isotropic turbulence DNS, and operates over it using a modified Physics-Informed Neural Network (PINN) \cite{raissi_physics-informed_2019}. PINNs have been used to solve inverse problems \cite{raissi_hidden_2020, shukla_physics-informed_2020} and to reconstruct turbulent flows out of measurements \cite{cai_flow_2021, clark_pressure_2022}. The Physics-Informed term regularizes the training by enforcing the residuals of the Navier-Stokes equation evaluated over the solution to be close to zero, and thus keeping the solution fluid-like. Our method is based on adding the aforementioned conditions as extra terms to the PINN loss function, thereby producing a resulting field that is close to the seed data and satisfies both the Navier-Stokes equations and the target conditions (up to a certain error).
The method is flexible and can be adapted to several kind of conditions. We also show two ways of scaling up the resolution of the generated states, one based on decomposing the spatio-temporal domain and running parallel PINNs, and the other based on nudging \cite{clark_di_leoni_inferring_2018, clark_di_leoni_synchronization_2020}, a synchronization-based DA technique which makes use of a classical numerical solver, with all its benefits and drawbacks.

The paper is organized as follows. In Section~\ref{sec:methods} we present the preparation method and the upscale procedure. In Section~\ref{sec:experiments} we describe the different experiments performed in this work. In Section~\ref{sec:results} we show the results obtained. Finally, in Section~\ref{sec:conclusions} we outline our conclusions and future lines of work.

\section{Methods}
\label{sec:methods}

\subsection{Modifying physics-informed neural networks}
\label{sec:modPINN}

The preparation method we present is based on modified PINNs \cite{raissi_physics-informed_2019}. In its original form, PINNs approximate solutions of partial differential equations. They take space and time coordinates as inputs, and output the desired fields. Their loss function is comprised of a standard $L^2$ norm of the output and the available data, the data term, and a regularization term composed of the residuals of the partial differential equations, the so-called physics term. During training the loss function is minimized, and the result is a regression on the data that satisfies the equations of motion. The architecture of the neural network itself is usually a standard fully-connected multilayer perceptron. For our purposes, we modified the PINN in two ways. The first is by adding a target term to the loss function which enforces the target constraint that we want to impose. Taking $(x,y,z)$ to be the three cartesian spatial coordinates, $\bm{u}^0$ the data available on the velocity field, and $\bm{u}=(u, v, w)$ the velocity field and $p$ the pressure outputted by the PINN, and using the incompressible Navier-Stokes equations with viscosity $\nu$ as our partial differential equations of choice, the total loss function then takes the form
\begin{align}
\label{eq:loss_general}
L_d  =  L_d + \lambda_p L_p + \lambda_t L_t,
\end{align}
where
\begin{align}
\label{eq:loss_data}
L_d  = \frac{1}{N_b}\sum_{\{i\}} \left(\bm{u}_i - \bm{u}^0_i \right)^2 ,
\end{align}
is the data term,
\begin{align}
\label{eq:loss_physics}
L_p  = \frac{1}{N_b}\sum_{\{i\}} \left[\left(
\frac{\partial \bm{u}_i}{\partial t}
+ \bm{u}_i\cdot \bm{\nabla} \bm{u}_i
+ \bm{\nabla} p_i - \nu \nabla^2 \bm{u}_i
\right)^2 + \left( \bm{\nabla} \cdot \bm{u} \right)^2 \right],
\end{align}
is the physics term, and $L_t$ is a problem-dependent target term. The subscript $i$ labels the point and time in which the fields are evaluated, i.e., $\bm{u}_i=\bm{u}(x_i, y_i, z_i, t_i)$. The whole set of points is separated into minibatches of size $N_b$, and the actual set of points $\{i\}$ that makes up each minibatch is picked at random out of the whole spatio-temporal domain where the problem is defined at each iteration. The hyperparameters $\lambda_p$ and $\lambda_t$ act as balancing terms between the different parts of the loss function. In typical neural network fashion, a PINN can then be trained using a mini-batch gradient-descent algorithm such as Adam.

The other modification we introduce is that we update the data term $\bm{u}^0$ after a certain number of iterations, and then continue with the training.
The initial data used when starting the training acts as a seed, and should come from a previously-performed simulation of the partial differential equations (PDEs), which in our case are the Navier-Stokes equations. Once the first training cycle is completed, the seed data is replaced by 
the output of the PINN at that stage. Thus, in each data update cycle, the data in the data term are closer to satisfying the target term in the loss function.
In DA parlance, the seed data would be the system state coming from a previous forecast, while the output of the whole training procedure would be the analysis.

\subsection{Upscaling the prepared fields}
\label{sec:upscaling}

Since the numerical study of turbulent flows requires high spatial resolution, we need to be able to increase the resolution of the PINN-generated state. This can be a challenge for neural networks, as computation of turbulent states at large Reynolds numbers usually require significant amounts of memory and computing power. One way of achieving this is by decomposing our spatio-temporal domain and running several PINNs in parallel, as per the C-PINN method \cite{karniadakis_extended_2020}. While we rely on this idea to increase the total time window used (more details on this below), it still presents considerable technical challenges when trying to upscale states to very high spatial resolutions.
Therefore, we present another way of upscaling the prepared fields based on the nudging technique \cite{clark_di_leoni_inferring_2018, clark_di_leoni_synchronization_2020}. Nudging works by adding a relaxation term to the r.h.s.~of the equations of motion that penalizes the evolved flow when it strays away from some given reference data, ${\bm{u}}_\text{ref}$. The resulting equations take the form
\begin{equation}
    \partial_t {\bm u} + ({\bm u}\cdot {\bm \nabla}) {\bm u} = -{\bm \nabla} p + \nu \nabla^2 {\bm u} - \alpha \mathcal{I}({\bm u} - {\bm u}_\text{ref}),
\label{eq:nudging}
\end{equation}
where the last term on the r.h.s.~corresponds to the nudging term, which penalizes the distance between the input reference data ${\bm u}_\text{ref}$ and ${\bm u}$. The amplitude of the nudging term is controlled with $\alpha$, and $\mathcal{I}$ is a filter operator which acts only where data is available.
In this work, $\mathcal{I}$ is a band-pass filter in Fourier space, which projects the spatial part of ${\bm u}({\bm x}, t)$ on the range of modes $[k_0, k_1]$ in which the reference field ${\bm u}_\text{ref}$ is known,
\begin{equation}
    \mathcal{I}{\bm u}({\bm x}, t) = \sum\limits_{k_0 \leq \lvert{\bm k}\lvert \leq k_1} \hat{\bm{u}}(\mathbf{k},t)~ \text{exp}(i{\bm k}\cdot {\bm x}).
\end{equation}

As a rule, the reference field (in our case, the output of the PINN) is also known in a given time interval $[0,T]$, with a time cadence $\tau$. The initial condition used to evolve Eq.~(\ref{eq:nudging}) corresponds to setting ${\bm u}(t= 0) = {\bm u}_\text{ref}(t=0)$. For the evolution in between successive observations of ${\bm  u}_\text{ref}$ (i.e., for time steps shorter than $\tau$), this field is linearly interpolated. Note that Eq.~(\ref{eq:nudging}) is evolved in time only in the interval $[0, T]$, for which ${\bm u}_\text{ref}$ is available.

Under this kind of setup, nudging has been shown to be able to reconstruct the flow at the scales where information is provided, while filling in the smaller scales with dynamics that must be compatible with the nudged scales even at high Reynolds number \cite{clark_di_leoni_inferring_2018, clark_di_leoni_synchronization_2020}. Thus, the upscaled field will reproduce the prepared field in the larger scales, and have dynamically consistent turbulent smalls scales. Finally, as to solve Eq.~\eqref{eq:nudging} a ``classical'' PDE solver must be used (e.g., a pseudospectral solver), we have the added advantages of having superior error convergence in the final states. To this end, we can make use of existing and already available highly-scalable parallel solvers \cite{mininni_hybrid_2011}.

Note the procedure discussed here is different from methods that use large-eddy models to generate the large scales in the flow, and use neural networks to fill in the missing information on the small scales of the flow \cite{Gamahara2017, Xie2020}. Instead, here the PINN is used to generate data compatible with the physics and with available large scale or statistical information (e.g., from observational data or experiments), and a PDE solver is then used to generate physically compatible small scale flow features.

\section{Experiments}
\label{sec:experiments}

We report on three separate experiments. In all three experiments the original seed field came from a homogeneous and isotropic forced DNS of the Navier-Stokes equation, performed with a resolution of $32^3$ grid points using a pseudospectral method with periodic boundary conditions, in a computational box of size $(2\pi {L_0})^3$ and in a time window of $T=3T_0$, where $L_0$ and $T_0$ are respectively the characteristic length and time scales of the flow, and amounting to a Reynolds number of $\mathrm{Re}= U_0 L_0/\nu=80$. The low spatial resolution is associated with limitations of the PINN, but also used to highlight the upscaling procedure. In light of the motivations for this work (cases in which observations or laboratory measurements are incomplete, e.g., with access to statistical information of only one field component), all preparations are performed over the $x$-component of the velocity. In the seed this component has zero mean, standard deviation $\sigma_0=0.5 U_0$, centralized third order moment $0.175 U_0$ and centralized fourth order moment $0.66 U_0$, where $U_0 = L_0/T_0$ is the flow characteristic velocity.

In Experiment 1 we use the method to impose a target mean profile on the $x$-component of the velocity field with the shape $u_0(y) = 0.1 U_0 \sin(y)$. The target term in the loss function given by Eq.~\eqref{eq:loss_general} then takes the form
\begin{equation}
L_t  = \left( \left[ \frac{1}{N_b}\sum_{\{i\}} u_i(y) \right] - u_0(y) \right)^2,
\label{eq:loss_1}
\end{equation}
where the minibatch subsets $\{i\}$ are all at different but fixed values of $y$.

In Experiment 2 we impose the value of the centralized third order moment of $u$ to be $s^3_0$ (which, equivalently, sets the skewness to be $s^3_0/\sigma^3_0$).
The target term takes the form
\begin{align}
\label{eq:loss_2}
L_t  =  \;\; &\left(\frac{1}{N_b}\sum_{\{i\}} u_i \right)^2 
\\
\nonumber
+&\left( \sqrt{\frac{1}{N_b}\sum_{\{i\}} u^2_i - \left(\frac{1}{N_b}\sum_{\{i\}} u_i \right)^2}  - \sigma_0 \right)^2 
\\
\nonumber
+&\left( {\frac{1}{N_b}\sum_{\{i\}} \left[ u_i - \frac{1}{N_b}\sum_{\{i\}} u_i \right]^3}  - s^3_0 \right)^2 ,
\end{align}
where the first and second terms are added to keep the mean and standard deviation from changing values. Note that we use the third order moment and not its cubic root in the loss function as this results in better convergence.

Finally, in Experiment 3 we impose the value of the fourth order moment to be $k^4_0=1$ (which, equivalently, sets the kurtosis to be $k^4_0/\sigma_0$). The target term takes the form
\begin{align}
\label{eq:loss_3}
L_t  =  \;\; &\left(\frac{1}{N_b}\sum_{\{i\}} u_i \right)^2 
\\
\nonumber
+&\left( \sqrt{\frac{1}{N_b}\sum_{\{i\}} u^2_i - \left(\frac{1}{N_b}\sum_{\{i\}} u_i \right)^2}  - \sigma_0 \right)^2 
\\
\nonumber
+&\left( {\frac{1}{N_b}\sum_{\{i\}} \left[ u_i - \frac{1}{N_b}\sum_{\{i\}} u_i \right]^4}  - k^4_0 \right)^2 .
\end{align}
While in principle the method works too if one tries to imposes the third and the fourth order moments, this can impose tension when training as the third order moment tires to skew the distribution while the fourth tries to symmetrize it. Therefore, due to this fact and the lack of a properly motivating case to include both moments, we decided to not include a fourth Experiment where both moments were modified.

The same neural network architecture was used in all three cases, a fully-connected multilayer perceptron with sines as activation functions, in practice a SIREN \cite{sitzmann_implicit_2020} with 8 hidden layers of 200 neurons each. The spatio-temporal domain of volume $(2\pi L_0)^3$ and time window length $T=3 T_0$ was split into four subdomains along the temporal dimension, so in each experiment actually four networks were trained with the final results concatenated at the end and treated as a whole. The balancing hyperparameters were chosen to be constant and with values $\lambda_p = 10^{-4}$ and $\lambda_t=1$, and the minibatches had $N_b=10000$. Following standard practices \cite{clark_pressure_2022} we added input and output normalization layers. We performed four data update cycles, for each cycle the networks were first optimized for $E_0$ epochs (where an epoch equals the number of iterations required to cover the whole dataset with mini-batch samples) using a learning rate of $5\times10^{-5}$, followed by $E_1$ epochs with a learning rate of $5\times10^{-6}$. In Experiment 1 $E_0=1000$ and $E_1=1000$, in Experiment 2 $E_0=500$ and $E_1=1000$, and in Experiment 3 $E_0=1000$ and $E_1=2000$.

Once training was completed, the resulting PINNs were evaluated on a uniform grid of $512^3$ spatial points every $\Delta t = 3.75\times 10^{-2}~ T_0$, resulting in a total of 80 snapshots spanning a time window of $T=3T_0$. These fields were used as the reference fields in the nudging-based upscaling procedure, as described in Sec.~\ref{sec:upscaling}. The nudging was performed with the same solver and boundary conditions used to generate the seed field. The band-pass filter $\mathcal{I}$ acted between  $k_0 = 1/L_0$ and $k_1 = 9/L_0$, which corresponds to the range of wave numbers contained in the original $32^3$ resolution seed field. The Reynolds number of the nudged simulation is or order $1000$.

\section{Results}
\label{sec:results}

\subsection{States generated by the PINN}
\label{sec:results}

\begin{figure}
    \centering
    \includegraphics[width=0.45\textwidth]{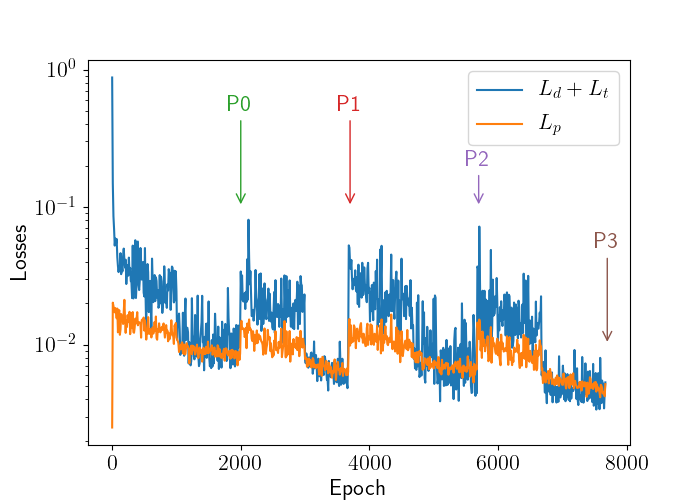}
    \includegraphics[width=0.45\textwidth]{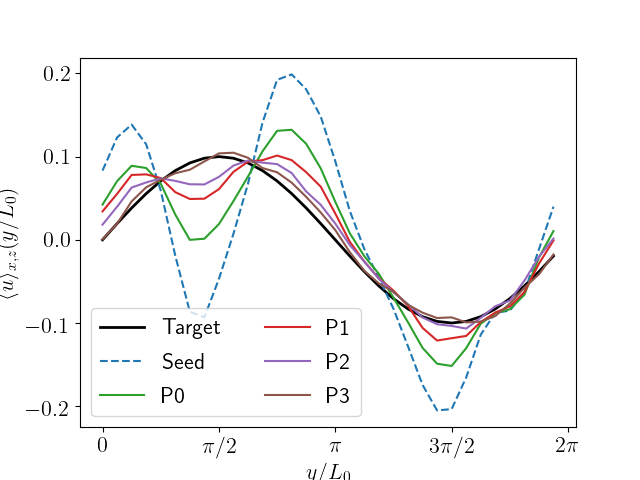}
    \caption{Results for the training of the PINN in Experiment 1: (a) Evolution of the $L_d + L_t$ and $L_p$ losses as a function of the training epoch. (b) Mean target profile, and mean $u$ profile at $t=1.125 T_0$ for the seed field and for the PINN-prepared field at different instants during training ($P0, P1, P2$, and $P3$ respectively, as marked in panel (a)).}
    \label{fig:pinn:exp1}
\end{figure}

\begin{figure}
    \centering
    \includegraphics[width=0.98\textwidth]{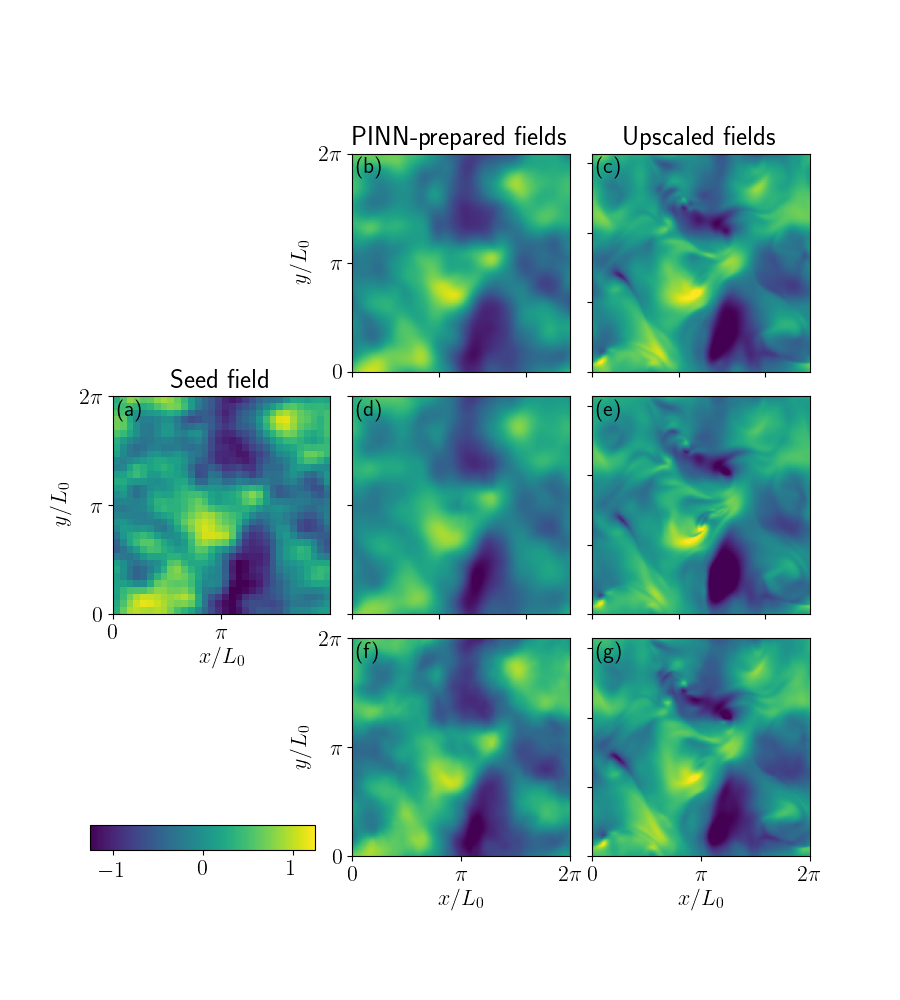}
    \caption{Visualizations of a two-dimensional slice of (a) the seed ($u$ velocity component), (b) the PINN output when imposing a mean velocity profile with (c) its corresponding nudging-upscaled field, (d) the PINN output with third order moment imposed and (e) upscaled field, and (f) the PINN output with fourth order moment imposed and (g) upscaled field. All slices correspond to time $t=1.125T_0$ and $z=\pi/L_0$.}
    \label{fig:vizes}
\end{figure}

We start by presenting results for Experiment 1. In figure~\ref{fig:pinn:exp1}(a) we show the evolution of the data plus target and physics parts of the loss function, while in panel (b) we show the target mean profile compared against the mean profile of the seed and those of the output at the end of each data cycle (labeled as $P0$ to $P3$, and indicated with arrows in panel (a)). Visualizations at $t=1.125T_0$ and $z=\pi/L_0$ of the seed and the PINN final prepared field, i.e., $P3$, are shown in Figure~\ref{fig:vizes}(a) and (b), respectively. Through the successive iterations and cycles, the preparation method is able to impose the mean profile while still resembling the original seed field and also complying with the Navier-Stokes equations (up to some error of order $5\times10^{-3}$ as per $L_p$ in Fig.~\ref{fig:pinn:exp1}(a)), thus retaining its fluid-like qualities.

\begin{figure}
    \centering
    \includegraphics[width=0.45\textwidth]{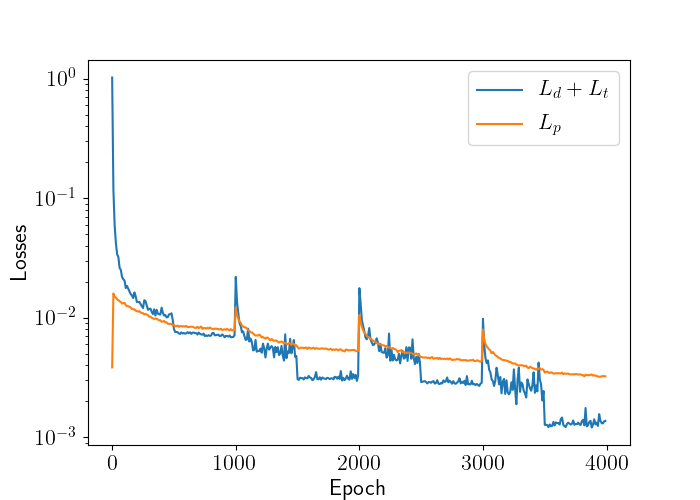}
    \includegraphics[width=0.45\textwidth]{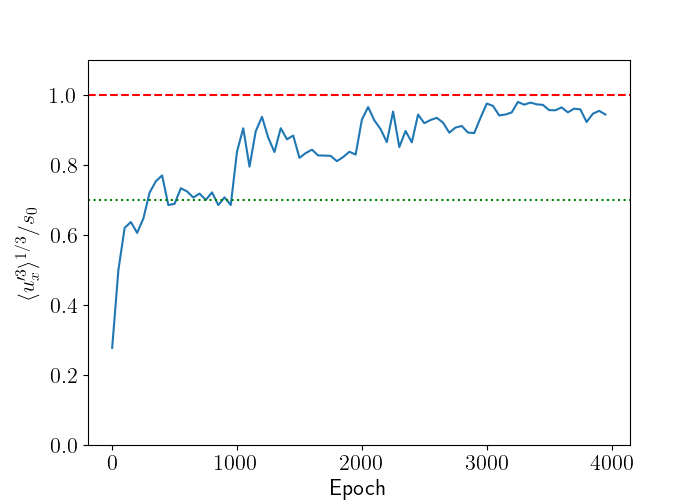}
    \caption{PINN results for Experiment 2: (a) Evolution of the $L_d + L_t$ and $L_p$ lesses as a function of the training epoch. (b) Evolution of the centralized third order moment normalized by the target $s_0$, as a function of the training epoch. The dashed red line indicates the target value, while the dotted green line indicates the value of the seed field.}
    \label{fig:pinn:exp2}
\end{figure}

Experiments 2 and 3 show similar results. In figure~\ref{fig:pinn:exp2}(a) we show the evolution of the different parts of the loss function for Experiment 2, while in panel (b) of the same figure we show the evolution of the centralized third order moment normalized by the target value $s_0$. The dashed red line indicates the target value we want to attain, while the dotted green line indicates the value of the original seed field. The prepared field third order moment converges to the desired value after two or three update cycles of the PINN. In figure~\ref{fig:vizes}(e) we show a visualization of the prepared field, which again shows a resemblance with the seed field, although a close inspection reveals differences with the result prepared in Experiment 1 in figure \ref{fig:vizes}(b).

\begin{figure}
    \centering
    \includegraphics[width=0.45\textwidth]{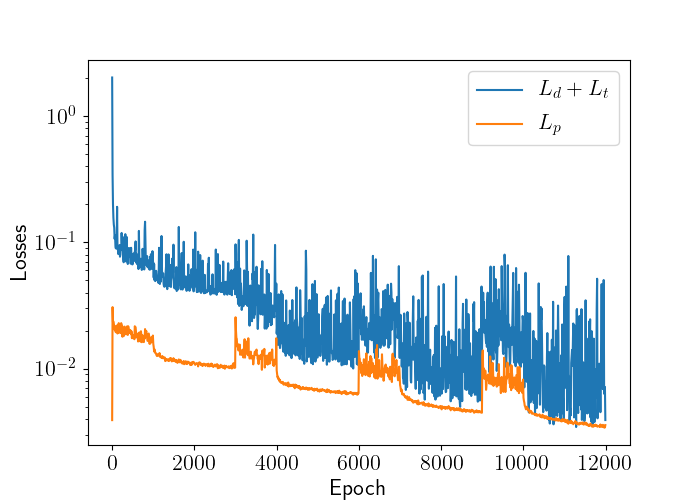}
    \includegraphics[width=0.45\textwidth]{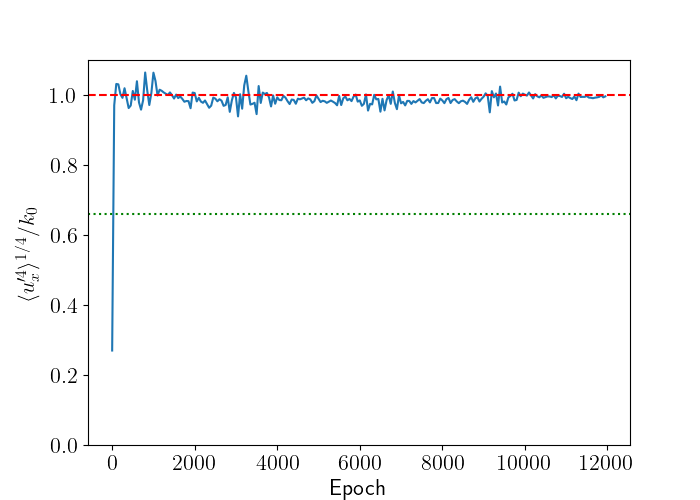}
    \caption{PINN results for Experiment 3: (a) Evolution of the $L_d + L_t$ and $L_p$ losses as a function of the training epoch. (b) Evolution of the centralized fourth order moment normalized by the target $k_0$, as a function of the training epoch. The dashed red line indicates the target value, while the dotted green line indicates the value of the seed field.}
    \label{fig:pinn:exp3}
\end{figure}

Finally, the results for Experiment 3 are shown in figure~\ref{fig:pinn:exp3}, where we plot the evolution of the loss function in panel (a), and the evolution of the centralized fourth order moment normalized by $k_0$ in panel (b). The dashed red and dotted green lines in panel (b) indicate the target and original values of the seed, respectively. Compared to Experiment 2, the fourth order moment converges much faster than the third order moment. Moreover, the value obtained after convergence fluctuate less after a few data update cycles.

\subsection{Upscaling}

\begin{figure}
    \centering
    \includegraphics[width=0.49\textwidth]{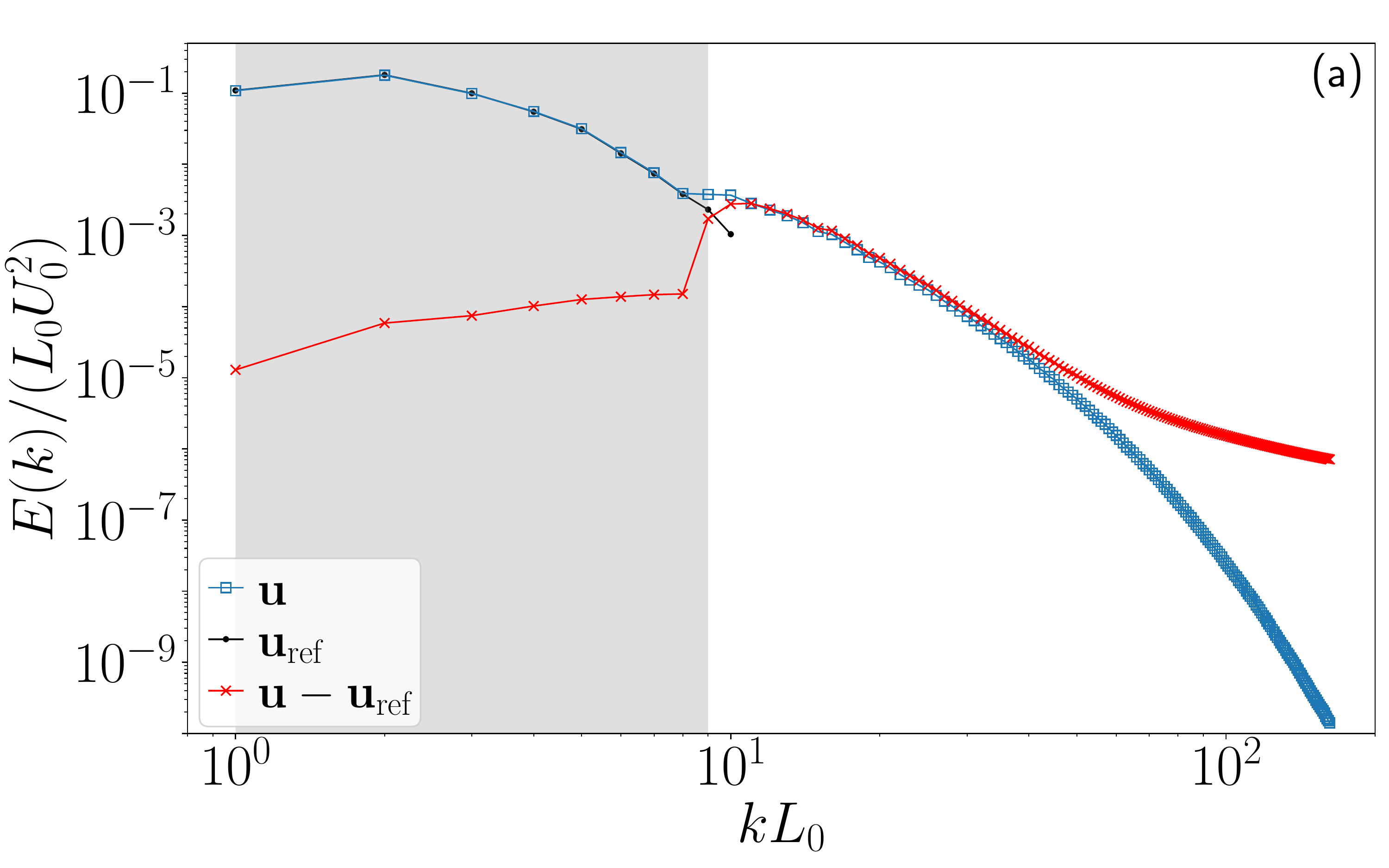}
    \includegraphics[width=0.49\textwidth]{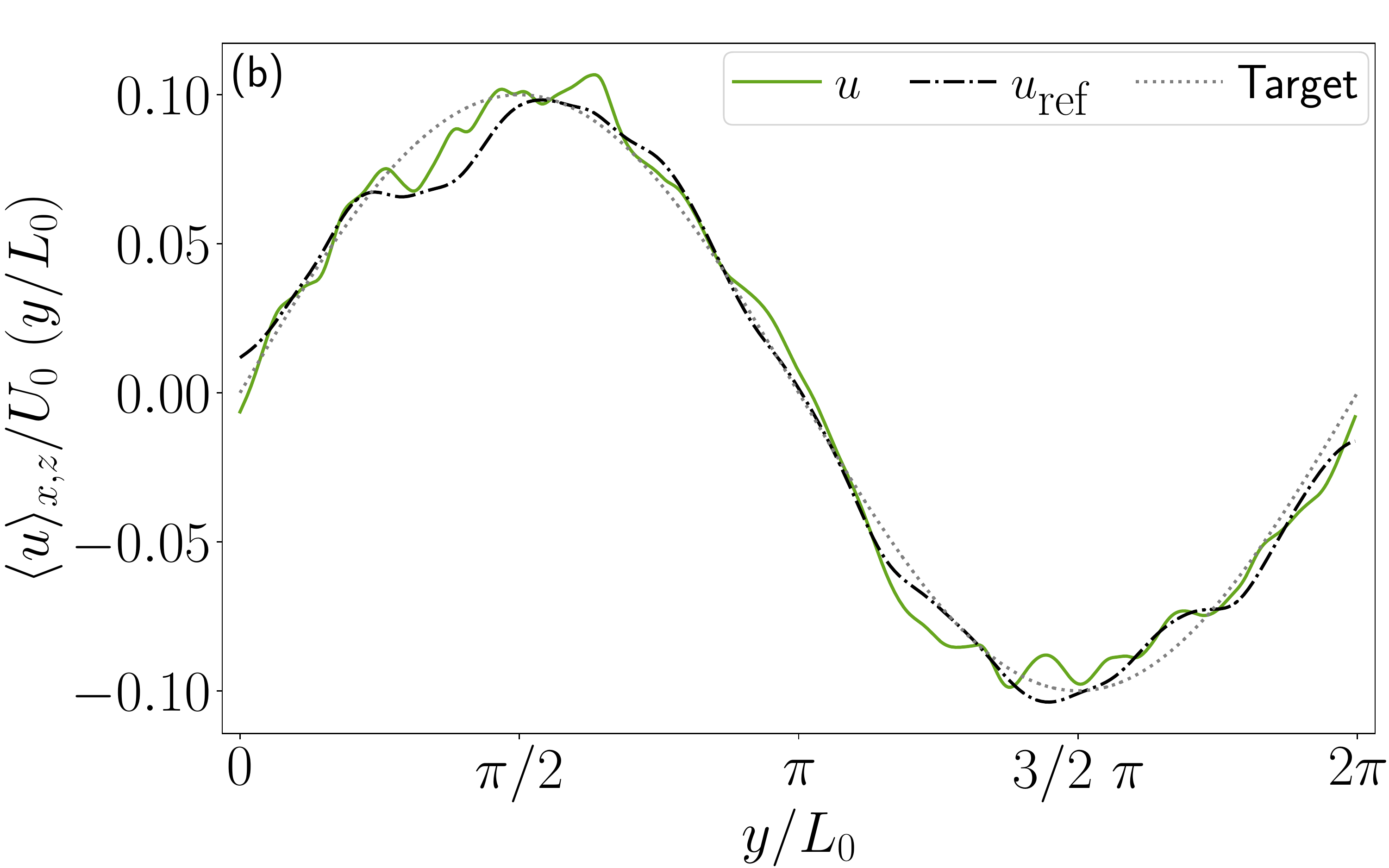}
    \hfill
    \caption{Nudging of Experiment 1. (a) Instantaneous energy spectra of the nudged field ${\bf u}$, the reference field ${\bf u}_\textrm{ref}$, and of the difference ${\bf u} - {\bf u}_\textrm{ref}$. All of the spectra were computed at a fixed time $t/T_0 = 1.125$. The shaded area corresponds to the Fourier modes in which the nudging is imposed.
    (b) Mean profile of the $x$-component of the nudged velocity field, as a function of $y/L_0$ at time $t/T_0 = 1.125$. The profile of the reference field and the target profile are shown for comparison.}
    \label{fig:nudging:exp1}
\end{figure}

\begin{figure}
    \centering
    \includegraphics[width=0.49\textwidth]{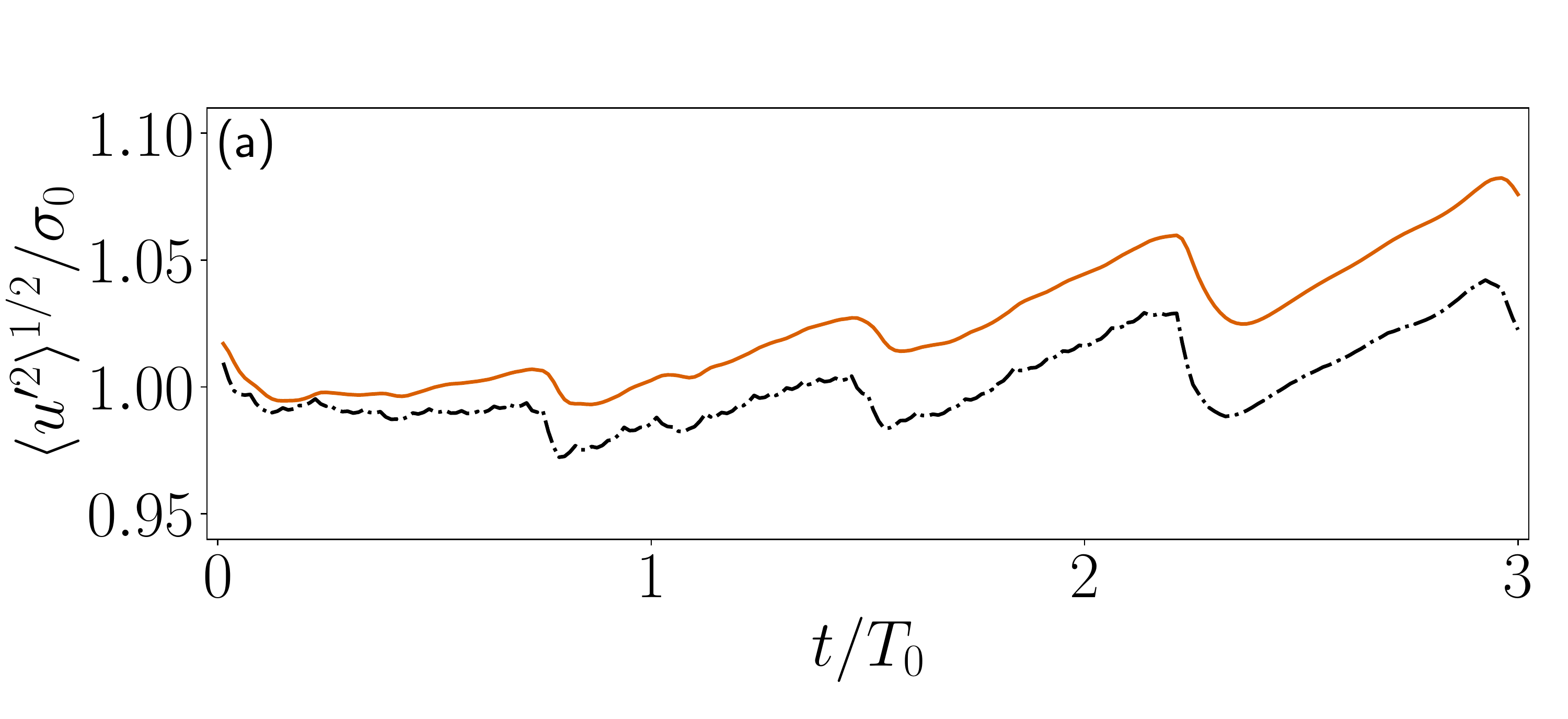}
        \includegraphics[width=0.49\textwidth]{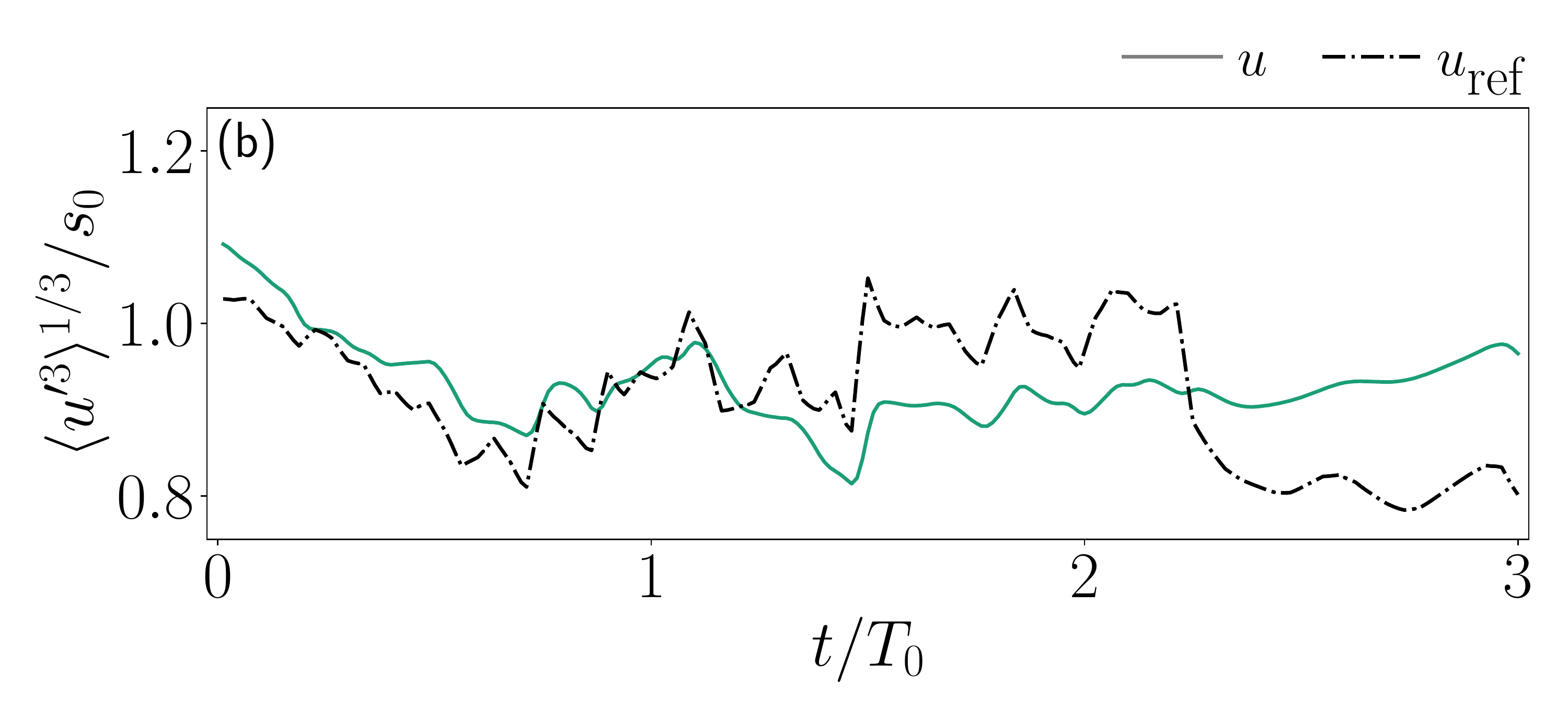}
    \hfill
    \caption{Nudging of Experiment 2. (a) Time evolution of the standard deviation of $u$ and $u_{\textrm{ref}}$, normalized by the target value $\sigma_0$. 
    (b) Time evolution of the centralized third order moment of $u$ and of $u_{\textrm{ref}}$, normalized by the target value $s_0$.}
    \label{fig:nudging:exp2}
\end{figure}

\begin{figure}
    \centering
    \includegraphics[width=0.49\textwidth]{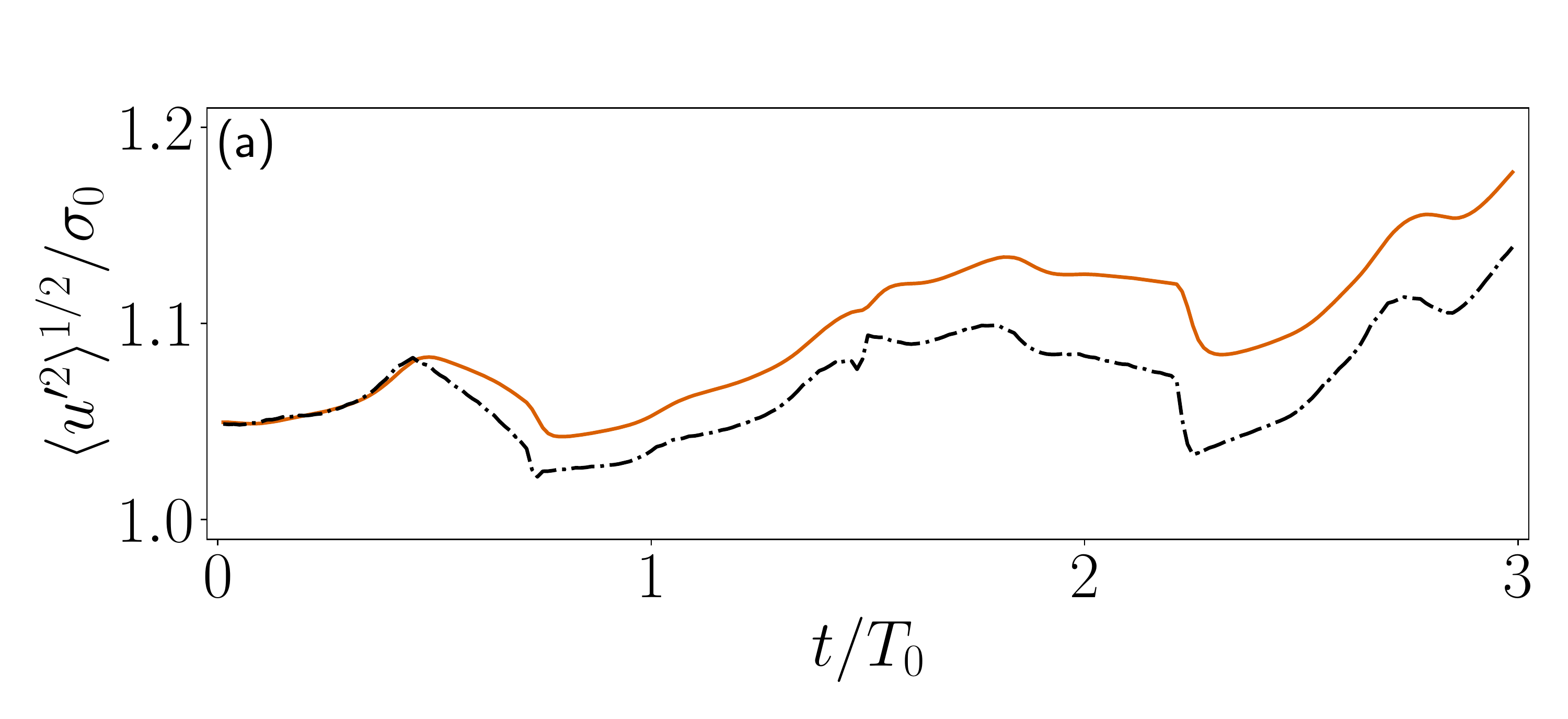}
    \includegraphics[width=0.49\textwidth]{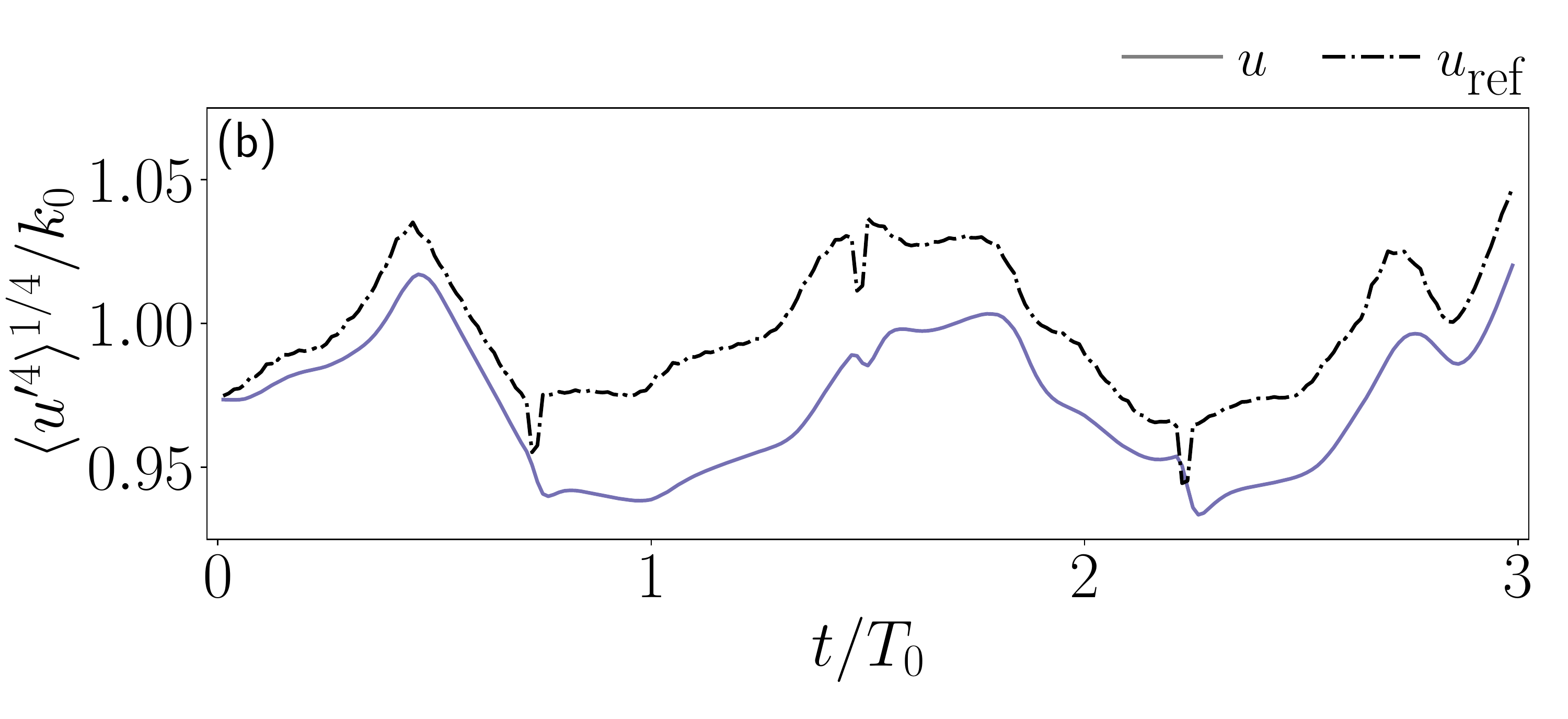}
    \hfill
    \caption{Nudging of Experiment 3. (a) Time evolution of the standard deviation of $u$ and $u_{\textrm{ref}}$, normalized by the target value $\sigma_0$. 
    (b) Time evolution of the centralized fourth order moment of $u$ and of $u_{\textrm{ref}}$, normalized by the target value $k_0$.}
    \label{fig:nudging:exp3}
\end{figure}

In this section we present the results of the nudging-based upscaling procedure. As explained above, the PINN-prepared fields obtained in the previous section are used as reference fields to nudge a simulation and generate upscaled versions (i.e., a version of the fields expected to have physically valid finer details and small scale features, while keeping the imposed conditions by the PINN).

Figure~\ref{fig:nudging:exp1}(a) shows the energy spectra of the nudged field ${\bf u}$, of the reference field ${\bf u}_\text{ref}$, and of the difference between both fields, ${\bf u} - {\bf u}_\text{ref}$, for Experiment 1.
The shaded region of the spectra indicates the range of nudged wave numbers (i.e., the range of scales in which the flow generated by the PINN has relevant information). In that region, the spectrum of the nudged simulation closely follows the spectrum of the reference field, while the spectrum of the difference of the two fields is several orders of magnitude smaller than any of the two fields, indicating that indeed the large scales of the nudged field are synchronized with the large scales of ${\bf u}_\text{ref}$. For $k L_0 \geq 10$ the nudged simulation fills in the missing scales, as energy cascades to smaller scales following the turbulent dynamics of the Navier-Stokes equations. In other words, the nudged field has a spectrum compatible with a turbulent flow, with an inertial range and a dissipative range. This is further confirmed in figure \ref{fig:vizes}(c), that shows a visualization of the nudged $u$ velocity field (i.e., the upscaled field), compared to the reference (i.e., the prepared field) in panel (b). The nudged field has finer structures while retaining the large scale characteristics of the prepared field.

Figure~\ref{fig:nudging:exp1}(b) shows the mean profile $\langle u\rangle_{x,z}$ (i.e., averaged over $x$ and $z$) as a function of $y/L_0$, for the nudged and reference velocity fields. The target mean profile is also shown for comparison. The nudged simulation is able to increase the scale separation of the prepared field (i.e., to create finer fluctuations) while maintaining the imposed mean profile. 

Upscaling experiments 2 and 3 leads to similar results. The resulting spectra are similar to those shown in figure~\ref{fig:nudging:exp1}(a) and as a result are not shown for the other experiments. As in the case of Experiment 1, the nudged fields have finer structures. This fact can be appreciated in the visualizations in figure \ref{fig:vizes}(e) and (g), respectively for experiments 2 and 3, where, again, the nudged simulations have smaller and finer structures than their reference fields.

In figure \ref{fig:nudging:exp2}(a) and (b) we show the time evolution of the standard deviation (normalized by the target $\sigma_0$) and the centralized third order moment (normalized by the target $s_0$) of the $x$-component of the nudged and reference velocity fields, respectively, for Experiment 2. And in figure \ref{fig:nudging:exp3}(a) and (b) we show the same comparison for the time evolution of the (normalized) standard deviation, and the centralized fourth order moment (normalized by the target $k_0$), respectively, for Experiment 3. In both cases the nudged simulations are able to synchronize to the reference data while also being able to capture the target statistic. The standard deviation of the nudged field departs slightly from the reference as the small scales are filled by the turbulent energy cascade, but on all cases the generated flows remain in the vicinity of the values imposed for the statistical moments.

\section{Conclusions}
\label{sec:conclusions}

The physics-informed neural network-powered method we presented is a flexible and powerful tool to prepare turbulent fields given a target statistical constraint. We showed three examples, one in which the target was a mean velocity profile, and two in which the targets were the values of the third and fourth order moments of the velocity field, respectively. In all cases, the method was able to prepare the field while retaining its fluid-like qualities. The method shows an example of the capabilities of deep learning in data assimilation problems.

Furthermore, we presented a nudging-based upscaling procedure which harnesses the power of already available and specialized numerical codes to increase the resolution of the prepared fields. Due to its specialized nature, this procedure is more efficient than heavily parallelizing the preparation problem and splitting it in multiple neural networks. The upscaling procedure is able to increase the scale separation of the prepared field (i.e., of generating physically compatible finer flow features), while maintaining the target statistics. The procedure also acts as a further reinforcement of the physics contained in the Navier-Stokes equations.
The combination of both procedures also provides an example of how neural networks can be combined with traditional numerical modeling of partial differential equations to overcome shortcomings in each separate procedure. 

As the method does not use observed data directly, but knowledge gathered from it, it can serve as a way to address problems with highly heterogeneous sources, such as atmospheric flows, where data is obtained from LIDAR measurements \cite{le_clainche_wind_2018}, satellite observations, and many more sources.
Besides applications in data assimilation, the proposed method can be useful in the classical problem of generation of initial conditions for turbulence simulations \cite{Rosales2005, Lavoie2007, gronskis_inflow_2013}, e.g., for the study of freely decaying homogeneous and isotropic turbulence, for grid generated turbulence in wind tunnels, or for turbulent flows with prescribed inflow boundary condition \cite{diMare2006, Perret2008, kim_deep_2020}.

\bibliography{sn-bibliography}


\begin{thebibliography}{27}
\ifx \bisbn   \undefined \def \bisbn  #1{ISBN #1}\fi
\ifx \binits  \undefined \def \binits#1{#1}\fi
\ifx \bauthor  \undefined \def \bauthor#1{#1}\fi
\ifx \batitle  \undefined \def \batitle#1{#1}\fi
\ifx \bjtitle  \undefined \def \bjtitle#1{#1}\fi
\ifx \bvolume  \undefined \def \bvolume#1{\textbf{#1}}\fi
\ifx \byear  \undefined \def \byear#1{#1}\fi
\ifx \bissue  \undefined \def \bissue#1{#1}\fi
\ifx \bfpage  \undefined \def \bfpage#1{#1}\fi
\ifx \blpage  \undefined \def \blpage #1{#1}\fi
\ifx \burl  \undefined \def \burl#1{\textsf{#1}}\fi
\ifx \doiurl  \undefined \def \doiurl#1{\url{https://doi.org/#1}}\fi
\ifx \betal  \undefined \def \betal{\textit{et al.}}\fi
\ifx \binstitute  \undefined \def \binstitute#1{#1}\fi
\ifx \binstitutionaled  \undefined \def \binstitutionaled#1{#1}\fi
\ifx \bctitle  \undefined \def \bctitle#1{#1}\fi
\ifx \beditor  \undefined \def \beditor#1{#1}\fi
\ifx \bpublisher  \undefined \def \bpublisher#1{#1}\fi
\ifx \bbtitle  \undefined \def \bbtitle#1{#1}\fi
\ifx \bedition  \undefined \def \bedition#1{#1}\fi
\ifx \bseriesno  \undefined \def \bseriesno#1{#1}\fi
\ifx \blocation  \undefined \def \blocation#1{#1}\fi
\ifx \bsertitle  \undefined \def \bsertitle#1{#1}\fi
\ifx \bsnm \undefined \def \bsnm#1{#1}\fi
\ifx \bsuffix \undefined \def \bsuffix#1{#1}\fi
\ifx \bparticle \undefined \def \bparticle#1{#1}\fi
\ifx \barticle \undefined \def \barticle#1{#1}\fi
\bibcommenthead
\ifx \bconfdate \undefined \def \bconfdate #1{#1}\fi
\ifx \botherref \undefined \def \botherref #1{#1}\fi
\ifx \url \undefined \def \url#1{\textsf{#1}}\fi
\ifx \bchapter \undefined \def \bchapter#1{#1}\fi
\ifx \bbook \undefined \def \bbook#1{#1}\fi
\ifx \bcomment \undefined \def \bcomment#1{#1}\fi
\ifx \oauthor \undefined \def \oauthor#1{#1}\fi
\ifx \citeauthoryear \undefined \def \citeauthoryear#1{#1}\fi
\ifx \endbibitem  \undefined \def \endbibitem {}\fi
\ifx \bconflocation  \undefined \def \bconflocation#1{#1}\fi
\ifx \arxivurl  \undefined \def \arxivurl#1{\textsf{#1}}\fi
\csname PreBibitemsHook\endcsname

\bibitem{mininni_hybrid_2011}
\begin{barticle}
\bauthor{\bsnm{Mininni}, \binits{P.D.}},
\bauthor{\bsnm{Rosenberg}, \binits{D.}},
\bauthor{\bsnm{Reddy}, \binits{R.}},
\bauthor{\bsnm{Pouquet}, \binits{A.}}:
\batitle{A hybrid {MPI}–{OpenMP} scheme for scalable parallel pseudospectral
  computations for fluid turbulence}.
\bjtitle{Parallel Computing}
\bvolume{37}(\bissue{6–7}),
\bfpage{316}--\blpage{326}
(\byear{2011}).
\doiurl{10.1016/j.parco.2011.05.004}.
Accessed 2014-08-07
\end{barticle}
\endbibitem

\bibitem{kraichnan_diffusion_1970}
\begin{barticle}
\bauthor{\bsnm{Kraichnan}, \binits{R.H.}}:
\batitle{Diffusion by a {Random} {Velocity} {Field}}.
\bjtitle{The Physics of Fluids}
\bvolume{13}(\bissue{1}),
\bfpage{22}--\blpage{31}
(\byear{1970}).
\doiurl{10.1063/1.1692799}.
\bcomment{Publisher: American Institute of Physics}.
Accessed 2022-08-17
\end{barticle}
\endbibitem

\bibitem{dhamankar_overview_2015}
\begin{bchapter}
\bauthor{\bsnm{Dhamankar}, \binits{N.S.}},
\bauthor{\bsnm{Blaisdell}, \binits{G.A.}},
\bauthor{\bsnm{Lyrintzis}, \binits{A.S.}}:
\bctitle{An {Overview} of {Turbulent} {Inflow} {Boundary} {Conditions} for
  {Large} {Eddy} {Simulations} ({Invited})}.
In: \bbtitle{22nd {AIAA} {Computational} {Fluid} {Dynamics} {Conference}}.
\bsertitle{{AIAA} {AVIATION} {Forum}}.
\bpublisher{American Institute of Aeronautics and Astronautics},
  \blocation{???}
(\byear{2015}).
\doiurl{10.2514/6.2015-3213}.
\burl{https://arc.aiaa.org/doi/10.2514/6.2015-3213}
Accessed 2022-08-17
\end{bchapter}
\endbibitem

\bibitem{wu_inflow_2017}
\begin{barticle}
\bauthor{\bsnm{Wu}, \binits{X.}}:
\batitle{Inflow {Turbulence} {Generation} {Methods}}.
\bjtitle{Annual Review of Fluid Mechanics}
\bvolume{49}(\bissue{1}),
\bfpage{23}--\blpage{49}
(\byear{2017}).
\doiurl{10.1146/annurev-fluid-010816-060322}.
\bcomment{\_eprint: https://doi.org/10.1146/annurev-fluid-010816-060322}.
Accessed 2022-08-17
\end{barticle}
\endbibitem

\bibitem{kalnay_atmospheric_2003}
\begin{bbook}
\bauthor{\bsnm{Kalnay}, \binits{E.}}:
\bbtitle{Atmospheric {Modeling}, {Data} {Assimilation} and {Predictability}}.
\bpublisher{Cambridge University Press}, \blocation{???}
(\byear{2003}).
\bcomment{Google-Books-ID: zx\_BakP2I5gC}
\end{bbook}
\endbibitem

\bibitem{Cotter2020}
\begin{barticle}
\bauthor{\bsnm{Cotter}, \binits{C.}},
\bauthor{\bsnm{Crisan}, \binits{D.}},
\bauthor{\bsnm{Holm}, \binits{D.}},
\bauthor{\bsnm{Pan}, \binits{W.}},
\bauthor{\bsnm{Shevchenko}, \binits{I.}}:
\batitle{Data assimilation for a quasi-geostrophic model with
  circulation-preserving stochastic transport noise}.
\bjtitle{Journal of Statistical Physics}
\bvolume{179},
\bfpage{1186}--\blpage{1221}
(\byear{2020})
\end{barticle}
\endbibitem

\bibitem{chantry_opportunities_2021}
\begin{barticle}
\bauthor{\bsnm{Chantry}, \binits{M.}},
\bauthor{\bsnm{Christensen}, \binits{H.}},
\bauthor{\bsnm{Dueben}, \binits{P.}},
\bauthor{\bsnm{Palmer}, \binits{T.}}:
\batitle{Opportunities and challenges for machine learning in weather and
  climate modelling: hard, medium and soft {AI}}.
\bjtitle{Philosophical Transactions of the Royal Society A: Mathematical,
  Physical and Engineering Sciences}
\bvolume{379}(\bissue{2194}),
\bfpage{20200083}
(\byear{2021}).
\doiurl{10.1098/rsta.2020.0083}.
\bcomment{Publisher: Royal Society}.
Accessed 2021-03-11
\end{barticle}
\endbibitem

\bibitem{foures_data-assimilation_2014}
\begin{barticle}
\bauthor{\bsnm{Foures}, \binits{D.P.G.}},
\bauthor{\bsnm{Dovetta}, \binits{N.}},
\bauthor{\bsnm{Sipp}, \binits{D.}},
\bauthor{\bsnm{Schmid}, \binits{P.J.}}:
\batitle{A data-assimilation method for {Reynolds}-averaged
  {Navier}–{Stokes}-driven mean flow reconstruction}.
\bjtitle{Journal of Fluid Mechanics}
\bvolume{759},
\bfpage{404}--\blpage{431}
(\byear{2014}).
\doiurl{10.1017/jfm.2014.566}.
\bcomment{Publisher: Cambridge University Press}.
Accessed 2022-08-18
\end{barticle}
\endbibitem

\bibitem{mons_ensemble-variational_2021}
\begin{barticle}
\bauthor{\bsnm{Mons}, \binits{V.}},
\bauthor{\bsnm{Du}, \binits{Y.}},
\bauthor{\bsnm{Zaki}, \binits{T.A.}}:
\batitle{Ensemble-variational assimilation of statistical data in large-eddy
  simulation}.
\bjtitle{Physical Review Fluids}
\bvolume{6}(\bissue{10}),
\bfpage{104607}
(\byear{2021}).
\doiurl{10.1103/PhysRevFluids.6.104607}.
\bcomment{Publisher: American Physical Society}.
Accessed 2022-08-10
\end{barticle}
\endbibitem

\bibitem{raissi_physics-informed_2019}
\begin{barticle}
\bauthor{\bsnm{Raissi}, \binits{M.}},
\bauthor{\bsnm{Perdikaris}, \binits{P.}},
\bauthor{\bsnm{Karniadakis}, \binits{G.E.}}:
\batitle{Physics-informed neural networks: {A} deep learning framework for
  solving forward and inverse problems involving nonlinear partial differential
  equations}.
\bjtitle{Journal of Computational Physics}
\bvolume{378},
\bfpage{686}--\blpage{707}
(\byear{2019}).
\doiurl{10.1016/j.jcp.2018.10.045}.
Accessed 2021-07-08
\end{barticle}
\endbibitem

\bibitem{raissi_hidden_2020}
\begin{barticle}
\bauthor{\bsnm{Raissi}, \binits{M.}},
\bauthor{\bsnm{Yazdani}, \binits{A.}},
\bauthor{\bsnm{Karniadakis}, \binits{G.E.}}:
\batitle{Hidden fluid mechanics: {Learning} velocity and pressure fields from
  flow visualizations}.
\bjtitle{Science}
\bvolume{367}(\bissue{6481}),
\bfpage{1026}--\blpage{1030}
(\byear{2020}).
\doiurl{10.1126/science.aaw4741}.
\bcomment{Publisher: American Association for the Advancement of Science
  Section: Report}.
Accessed 2020-04-06
\end{barticle}
\endbibitem

\bibitem{shukla_physics-informed_2020}
\begin{barticle}
\bauthor{\bsnm{Shukla}, \binits{K.}},
\bauthor{\bsnm{Clark Di~Leoni}, \binits{P.}},
\bauthor{\bsnm{Blackshire}, \binits{J.}},
\bauthor{\bsnm{Sparkman}, \binits{D.}},
\bauthor{\bsnm{Karniadakis}, \binits{G.E.}}:
\batitle{Physics-{Informed} {Neural} {Network} for {Ultrasound}
  {Nondestructive} {Quantification} of {Surface} {Breaking} {Cracks}}.
\bjtitle{Journal of Nondestructive Evaluation}
\bvolume{39}(\bissue{3}),
\bfpage{61}
(\byear{2020}).
\doiurl{10.1007/s10921-020-00705-1}.
Accessed 2020-08-05
\end{barticle}
\endbibitem

\bibitem{cai_flow_2021}
\begin{botherref}
\oauthor{\bsnm{Cai}, \binits{S.}},
\oauthor{\bsnm{Wang}, \binits{Z.}},
\oauthor{\bsnm{Fuest}, \binits{F.}},
\oauthor{\bsnm{Jeon}, \binits{Y.J.}},
\oauthor{\bsnm{Gray}, \binits{C.}},
\oauthor{\bsnm{Karniadakis}, \binits{G.E.}}:
Flow over an espresso cup: inferring 3-{D} velocity and pressure fields from
  tomographic background oriented {Schlieren} via physics-informed neural
  networks.
Journal of Fluid Mechanics
\textbf{915}
(2021).
\doiurl{10.1017/jfm.2021.135}.
Publisher: Cambridge University Press
\end{botherref}
\endbibitem

\bibitem{clark_pressure_2022}
\begin{botherref}
\oauthor{\bsnm{Clark Di~Leoni}, \binits{P.}},
\oauthor{\bsnm{Agarwal}, \binits{K.}},
\oauthor{\bsnm{Zaki}, \binits{T.}},
\oauthor{\bsnm{Meneveau}, \binits{C.}},
\oauthor{\bsnm{Katz}, \binits{J.}}:
Pressure pinns.
In Preparation
(2021).
Publisher: Cambridge University Press
\end{botherref}
\endbibitem

\bibitem{clark_di_leoni_inferring_2018}
\begin{barticle}
\bauthor{\bsnm{Clark Di~Leoni}, \binits{P.}},
\bauthor{\bsnm{Mazzino}, \binits{A.}},
\bauthor{\bsnm{Biferale}, \binits{L.}}:
\batitle{Inferring flow parameters and turbulent configuration with
  physics-informed data assimilation and spectral nudging}.
\bjtitle{Physical Review Fluids}
\bvolume{3}(\bissue{10}),
\bfpage{104604}
(\byear{2018}).
\doiurl{10.1103/PhysRevFluids.3.104604}.
Accessed 2018-12-14
\end{barticle}
\endbibitem

\bibitem{clark_di_leoni_synchronization_2020}
\begin{barticle}
\bauthor{\bsnm{Clark Di~Leoni}, \binits{P.}},
\bauthor{\bsnm{Mazzino}, \binits{A.}},
\bauthor{\bsnm{Biferale}, \binits{L.}}:
\batitle{Synchronization to {Big} {Data}: {Nudging} the {Navier}-{Stokes}
  {Equations} for {Data} {Assimilation} of {Turbulent} {Flows}}.
\bjtitle{Physical Review X}
\bvolume{10}(\bissue{1}),
\bfpage{011023}
(\byear{2020}).
\doiurl{10.1103/PhysRevX.10.011023}.
\bcomment{Publisher: American Physical Society}.
Accessed 2020-02-28
\end{barticle}
\endbibitem

\bibitem{karniadakis_extended_2020}
\begin{barticle}
\bauthor{\bsnm{Karniadakis}, \binits{A.D.J..G.E.}}:
\batitle{Extended {Physics}-{Informed} {Neural} {Networks} ({XPINNs}): {A}
  {Generalized} {Space}-{Time} {Domain} {Decomposition} {Based} {Deep}
  {Learning} {Framework} for {Nonlinear} {Partial} {Differential} {Equations}}.
\bjtitle{Communications in Computational Physics}
\bvolume{28}(\bissue{5}),
\bfpage{2002}--\blpage{2041}
(\byear{2020}).
\doiurl{10.4208/cicp.OA-2020-0164}.
Accessed 2022-01-26
\end{barticle}
\endbibitem

\bibitem{Gamahara2017}
\begin{barticle}
\bauthor{\bsnm{Gamahara}, \binits{M.}},
\bauthor{\bsnm{Hattori}, \binits{Y.}}:
\batitle{Searching for turbulence models by artificial neural network}.
\bjtitle{Physical Review Fluids}
\bvolume{2},
\bfpage{054604}
(\byear{2017})
\end{barticle}
\endbibitem

\bibitem{Xie2020}
\begin{barticle}
\bauthor{\bsnm{Xie}, \binits{C.}},
\bauthor{\bsnm{Wang}, \binits{J.}},
\bauthor{\bsnm{Weinan}, \binits{E.}}:
\batitle{Modeling subgrid-scale forces by spatial artificial neural networks in
  large eddy simulation of turbulence}.
\bjtitle{Physical Review Fluids}
\bvolume{5},
\bfpage{054606}
(\byear{2020})
\end{barticle}
\endbibitem

\bibitem{sitzmann_implicit_2020}
\begin{botherref}
\oauthor{\bsnm{Sitzmann}, \binits{V.}},
\oauthor{\bsnm{Martel}, \binits{J.N.P.}},
\oauthor{\bsnm{Bergman}, \binits{A.W.}},
\oauthor{\bsnm{Lindell}, \binits{D.B.}},
\oauthor{\bsnm{Wetzstein}, \binits{G.}}:
Implicit {Neural} {Representations} with {Periodic} {Activation} {Functions}.
arXiv:2006.09661 [cs, eess]
(2020).
arXiv: 2006.09661.
Accessed 2022-04-19
\end{botherref}
\endbibitem

\bibitem{le_clainche_wind_2018}
\begin{barticle}
\bauthor{\bsnm{Le~Clainche}, \binits{S.}},
\bauthor{\bsnm{Lorente}, \binits{L.S.}},
\bauthor{\bsnm{Vega}, \binits{J.M.}}:
\batitle{Wind {Predictions} {Upstream} {Wind} {Turbines} from a {LiDAR}
  {Database}}.
\bjtitle{Energies}
\bvolume{11}(\bissue{3}),
\bfpage{543}
(\byear{2018}).
\doiurl{10.3390/en11030543}.
\bcomment{Number: 3 Publisher: Multidisciplinary Digital Publishing Institute}.
Accessed 2022-09-08
\end{barticle}
\endbibitem

\bibitem{Rosales2005}
\begin{barticle}
\bauthor{\bsnm{Rosales}, \binits{C.}},
\bauthor{\bsnm{Meneveau}, \binits{C.}}:
\batitle{Linear forcing in numerical simulations of isotropic turbulence:
  Physical space implementations and convergence properties}.
\bjtitle{Physics of fluids}
\bvolume{17},
\bfpage{095106}
(\byear{2005})
\end{barticle}
\endbibitem

\bibitem{Lavoie2007}
\begin{barticle}
\bauthor{\bsnm{Lavoie}, \binits{P.}},
\bauthor{\bsnm{Djenidi}, \binits{L.}},
\bauthor{\bsnm{Antonia}, \binits{R.}}:
\batitle{Effects of initial conditions in decaying turbulence generated by
  passive grids}.
\bjtitle{Journal of Fluid Mechanics}
\bvolume{585},
\bfpage{395}--\blpage{420}
(\byear{2007})
\end{barticle}
\endbibitem

\bibitem{gronskis_inflow_2013}
\begin{barticle}
\bauthor{\bsnm{Gronskis}, \binits{A.}},
\bauthor{\bsnm{Heitz}, \binits{D.}},
\bauthor{\bsnm{Mémin}, \binits{E.}}:
\batitle{Inflow and initial conditions for direct numerical simulation based on
  adjoint data assimilation}.
\bjtitle{Journal of Computational Physics}
\bvolume{242},
\bfpage{480}--\blpage{497}
(\byear{2013}).
\doiurl{10.1016/j.jcp.2013.01.051}.
Accessed 2022-08-11
\end{barticle}
\endbibitem

\bibitem{diMare2006}
\begin{barticle}
\bauthor{\bparticle{di} \bsnm{Mare}, \binits{L.}},
\bauthor{\bsnm{Klein}, \binits{M.}},
\bauthor{\bsnm{Jones}, \binits{W.P.}},
\bauthor{\bsnm{Janicka}, \binits{J.}}:
\batitle{Synthetic turbulence inflow conditions for large-eddy simulation}.
\bjtitle{Physics of Fluids}
\bvolume{18}(\bissue{2}),
\bfpage{025107}
(\byear{2006}).
\doiurl{10.1063/1.2130744}
\end{barticle}
\endbibitem

\bibitem{Perret2008}
\begin{barticle}
\bauthor{\bsnm{Perret}, \binits{L.}},
\bauthor{\bsnm{Delville}, \binits{J.}},
\bauthor{\bsnm{Manceau}, \binits{R.}},
\bauthor{\bsnm{Bonnet}, \binits{J.-P.}}:
\batitle{Turbulent inflow conditions for large-eddy simulation based on
  low-order empirical model}.
\bjtitle{Physics of Fluids}
\bvolume{20}(\bissue{7}),
\bfpage{075107}
(\byear{2008}).
\doiurl{10.1063/1.2957019}
\end{barticle}
\endbibitem

\bibitem{kim_deep_2020}
\begin{barticle}
\bauthor{\bsnm{Kim}, \binits{J.}},
\bauthor{\bsnm{Lee}, \binits{C.}}:
\batitle{Deep unsupervised learning of turbulence for inflow generation at
  various {Reynolds} numbers}.
\bjtitle{Journal of Computational Physics}
\bvolume{406},
\bfpage{109216}
(\byear{2020}).
\doiurl{10.1016/j.jcp.2019.109216}.
Accessed 2022-09-08
\end{barticle}
\endbibitem

\end{thebibliography}

\end{document}